# How Light Shapes Memory: Beta Synchrony in the Temporal-Parietal Cortex Predicts Cognitive Ergonomics for BCI Applications

**Jiajia Li** [1,2], **Tian Guo** [1], **Fan Li** [1], **Huichao Ding** [2,3], **Guozheng Xu** [2,3] * and **Jian Song** [2,3] *

[1] Department of Artificial Intelligence and Automation, School of Information and Control Engineering, Xi'an University of Architecture and Technology, Xi'an, 710055, China
[2] Department of Neurosurgery, PLA General Hospital of Central Theater Command, Wuhan, 430070, China
[3] The First School of Clinical Medicine, Southern Medical University, Guangzhou 510515, China
* Correspondence: songjian0505@smu.edu.cn(JS); xu-gz@163.com (GX)

**Abstract**

Working memory is a promising paradigm for assessing cognitive ergonomics of brain states in brain-computer interfaces (BCIs). This study decodes these states—with a focus on environmental illumination effects—via two distinct working memory tasks (Recall and Sequence) for mixed-recognition analysis. Leveraging nonlinear patterns in brain connectivity, we propose an innovative framework: multi-regional dynamic interplay patterns based on beta phase synchrony dynamics, to identify low-dimensional EEG regions (prefrontal, temporal, parietal) for state recognition. Based on nonlinear phase map analysis of the above three brain regions using beta-phase connectivity, we found that: (1) Temporal-parietal phase clustering outperforms other regional combinations in distinguishing memory states; (2) Illumination-enhanced environments optimize temporoparietal balance; (3) Machine learning confirms temporal-parietal synchrony as the dominant cross-task classification feature. These results provide a precise prediction algorithm, facilitating a low-dimensional system using temporal and parietal EEG channels—with practical value for real-time cognitive ergonomics assessment in BCIs and optimized human-machine interaction.

## 1. Introduction

Working memory (WM), a core component of the human cognitive system, is indispensable in daily and occupational activities—serving as a critical cognitive support system across contexts from construction sites to laboratories [1]. As a dynamic, adaptive brain system, it adjusts to external environmental changes or task difficulty to achieve self-optimization [2]. Thus, clarifying the patterns of adaptive memory states and their neurocognitive mechanisms holds significant theoretical and practical value, particularly for measuring and regulating memory capacity in extreme environments (e.g., real-time monitoring of astronauts' memory states to optimize task scheduling and resource allocation [3]).As a form of short-term memory, WM enables the brain to process ongoing tasks, facilitating perception and decision-making related to visual, auditory, and motor information through dynamic cortico-network mechanisms [[4,5]]. Since Baddeley and Hitch proposed the WM model [6], research has shown the prefrontal cortex (PFC) mediates both bottom-up perceptual processing attention formation and top-down regulation of posterior cortices [7,8,9], while coordinating with hippocampal activity to enhance memory retrieval efficiency [10]. However, complex cognitive processing relies on multi-regional cooperation: the PFC acts as a central control network, with temporal cortex (TC) beta/gamma networks activating alongside it [11,12], and parietal cortex (PC) playing a pivotal role in attention formation and memory retrieval [13,14]. These findings highlight the criticality of PFC-TC-PC coordination for effective WM function.

Current research has primarily focused on the static roles of temporal and parietal cortices in supporting the PFC during memory tasks. However, emerging evidence indicates memory states are plastic, dynamically adapting to environmental factors (e.g., lighting [15,16]) and task demands [17,18,19]—for example, increased memory load activates additional neural modules to maintain efficiency [20,21]. Yet, how the brain dynamically adjusts network structure via WM's adaptive mechanisms to meet cognitive and environmental demands remains poorly understood; elucidating this could deepen insights into cortical network homeostasis and dynamic memory regulation beyond static observations.

To address this, our study innovates in experimental design and methodology. Experimentally, we developed a multi-task adaptive WM paradigm integrating two task difficulty levels [22] and two lighting conditions [23], yielding four experimental combinations. This design adjusts WM load and processing efficiency to induce plastic changes and explore dynamic network characteristics [24]. Methodologically, we adopted a dynamic perspective, focusing on beta-band phase synchronization across parietal, temporal, and prefrontal regions to examine multi-regional dynamical patterns under different memory states. We proposed a physical explanation for multi-variable brain balance and used machine learning to compare classification accuracy across regional combinations (temporal-parietal, temporal-prefrontal, parietal-prefrontal, and multi-regional). These efforts aim to reveal dynamic mechanisms of critical cortical circuits underlying adaptive memory adjustments, advancing understanding of how brain networks optimize memory function through dynamic regulation.

## 2. Materials and Methods

*2.1. Experimental Environment*

The experimental site for this study was established in the study room of the school library, a space with dimensions of L5.0×W4.5×H3.0m. This room is isolated from the collective learning area and features an independently adjustable air conditioning system. The lighting system in the experimental environment includes dimming LED lighting fixtures installed in the study area. These fixtures, mounted using a recessed installation method, are suspended directly above the experimental tables. They provide a task plane illuminance of 0.7 meters above the desk surface, meeting the illuminance requirements for fine visual tasks specified in the lighting standards.

*2.2. Experimental Conditions*

To investigate working efficiency under different illuminance levels, we established two experimental lighting conditions: standard office illuminance at 300 lux and high-intensity illuminance at 1000 lux, based on previous research [25]. The laboratory was equipped with blackout curtains to maintain a stable background illuminance level. To ensure participants' comfort and minimize environmental interference, the following parameters were controlled: the indoor thermal environment was maintained at 26°C using an air conditioning system, and the indoor noise level was set to meet the requirements of a Class 0 acoustic standard (limit: 50 dB during daytime hours), providing standardized experimental conditions for cognitive effect measurements. The cognitive tasks were programmed using E-Prime 3.0[26] and presented on a 14-inch laptop screen. Before the experiment, the display parameters (such as brightness and contrast) were calibrated to eliminate potential differences in device performance. During the test, natural light interference was blocked by closing the blackout curtains, and precise control of the ambient lighting was achieved using LED devices. The test computer screen was placed

50 centimeters away from the participant. Participants completed between 50 to 100 task blocks across multiple test sessions.

*2.3. Participants*

In this study, 22 healthy male participants were recruited, all of whom are right-handed (mean age 24 years). The research protocol was approved by the Ethics Committee of the General Hospital of Chinese PLA Central Theater Command (reference number: [2020]041-1). Informed consent was obtained from all participants prior to the experiment. All participants were proficient in basic computer operations and possessed bilingual cognitive abilities in Chinese and English. They were free from eye diseases, neurological disorders, or other conditions that could influence the experimental results. In accordance with the experimental guidelines, participants were required to follow specific rules during the 24 hours preceding the experiment: maintain adequate sleep, refrain from consuming alcohol or other substances that could affect the nervous system, and ensure their scalp was clean. During the experiment, participants were required to remain focused, avoid speaking, and minimize significant head movements.

*2.4. Experimental Paradigms and Behavioral Statistics*

**Recall Paradigm:**

The Recall paradigm is primarily employed to examine the cognitive processing characteristics in participants during the information encoding and retrieval stages. It measures memory effects by having participants recall information they previously learned. In free recall, participants are initially presented with a learning list of 10 Chinese-English word pairs, which lack strong semantic associations. Afterwards, from a list containing additional words, participants must recall the words they saw earlier.

**Sequence Memory Paradigm：**

The Sequence Memory paradigm is an experimental approach used to study the storage and retrieval of sequential information in memory. It requires participants not only to remember the items themselves but also to recall them in their original order. In this task, participants are presented with 8 groups of short sentences, with the final character of each sentence serving as the target item to be remembered. They are required to remember the final character of each sentence and recall them in the correct sequence.

The experimental procedure, as illustrated in Figure 2A-C, examined the effect of light intensity on learning and memory by controlling ambient illuminance at two levels: 300 lx and 1000 lx. Based on neuroscientific theories suggesting that light environments influence cognitive processing, we hypothesized that increasing ambient illuminance would enhance working memory encoding efficiency and executive control functions. The experimental task utilizes a dual-dimensional evaluation system, comprising accuracy (ACC) and reaction time (RT). ACC indicates the number of successful recalls, reflecting the accuracy of cognitive processing, while RT measures the time taken to complete the task, indicating processing efficiency. In this experiment, ACC and RT for the Recall task were recorded using E-Prime. For the Sequence Memory task, ACC (defined as the number of characters recalled correctly in sequence) was calculated using a specific formula, defined as follows:

$$ACC = \frac{TA}{SA} \times 100\% \tag{1}$$

In Eq. (1), ACC represents accuracy, TA is the number of correctly recalled characters, and SA is the total number of characters presented.

As shown in Figure 1, the subjective results reveal that with the increase of illuminance, the brightness perception dimension exerts a significant main effect; meanwhile, visual satisfaction and spatial spaciousness are enhanced. Figure 2D and E display a comparative behavioral analysis of the two cognitive tasks under the two illuminance conditions (300 lx and 1000 lx). The experimental data indicate that the high-illuminance group showed significant advantages in both tasks. Specifically, under high-illuminance conditions, the Recall task demonstrated a 5.76% improvement in ACC and a 13.3% reduction in RT, while the Sequence Memory task showed a 13.7% improvement in ACC. These findings suggest that increased illuminance enhances the brain's information processing efficiency, leading to faster task completion and higher ACC, thus supporting the positive effect of optimized lighting on cognitive functions.

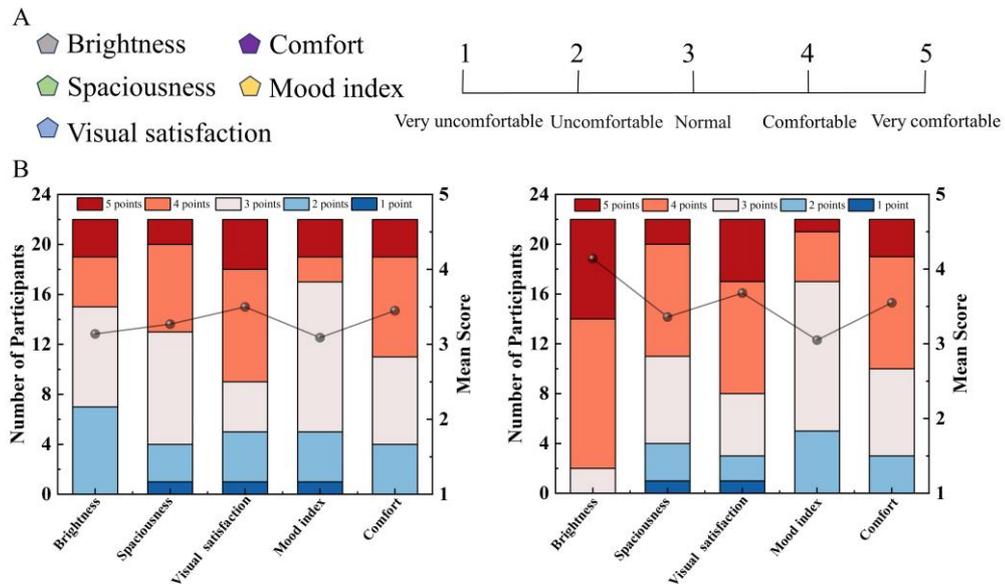

**Figure 1.** Statistics of subjective evaluation indicators. A. Likert Scale: This scale comprises a series of statement items, each offering five response options, scored on a scale of 1 to 5; B. Subjective Evaluation Statistics: Left panel: 300 lx results distribution; right panel: 1000 lx. Black line: mean of each subjective indicator.

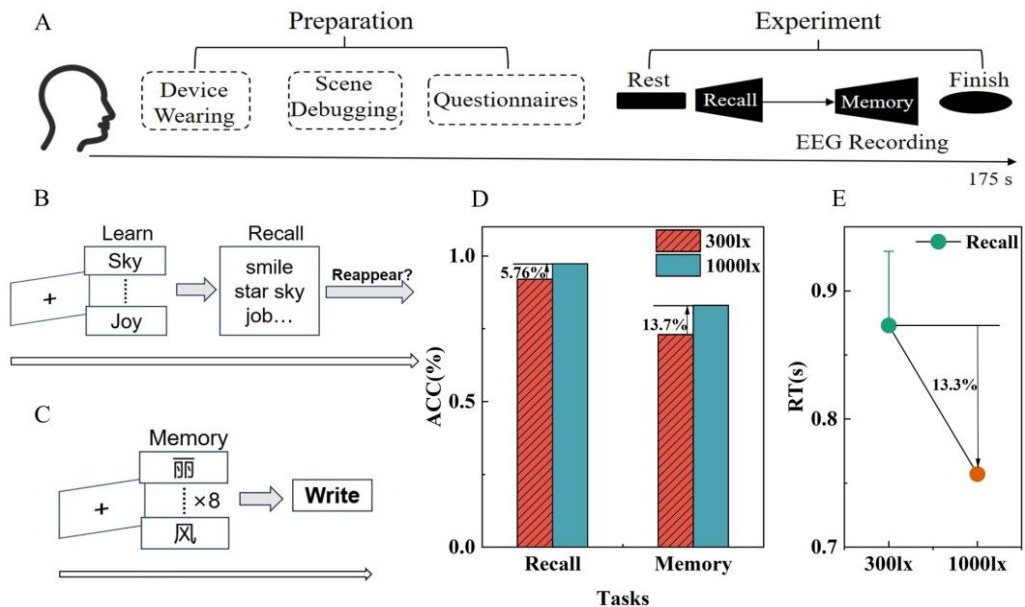

**Figure 2.** Working memory task paradigms and the behavioral performance statistics. A. Overall experimental flowchart. B. Flowchart of the Recall paradigm cognitive task; C. Flowchart of the Sequence memory paradigm cognitive task; D. Behavioral ACC statistics under different lighting conditions; E. Behavioral RT statistics under different lighting conditions.

Statistical results consistently indicate ACC under 1000 lx illuminance keep significantly higher than that under 300 lx conditions (p < 0.05). Differences in task accuracy (ACC) and reaction time (RT) among the 22 participants are presented in Tables 1 and 2, respectively.

Table 1 The Correlation Analysis of Accuracy Rates in Cognitive Task Tests

| Task Testing | Indoor lighting | Mean1 | Mean2 | Mean difference | t | P-value |
|---|---|---|---|---|---|---|
| Recall | 300lx vs 1000lx | 0.92 | 0.973 | -0.052 | -3.429 | 0.003*** |
| Memory | 300lx vs 1000lx | 0.73 | 0.83 | -0.102 | -2.144 | 0.044** |

Note：***, **, and * represent the significance levels of 1%, 5%, and 10% respectively.

Table 2 The Correlation Analysis of Reaction time in Cognitive Task Tests

| Task Testing | Indoor lighting | Mean1 | Mean2 | Mean difference | t | P-value |
|---|---|---|---|---|---|---|
| Recall | 300lx vs 1000lx | 0.815 | 0.757 | 0.058 | 2.095 | 0.049** |

Note：***, **, and * represent the significance levels of 1%, 5%, and 10% respectively.

*2.5. Analysis Algorithms*

We investigated working memory (WM) adaptability in brain neural responses using EEG, combining multi-regional phase synchronization analysis in beta-band oscillations with machine learning algorithms. EEG data preprocessing involved extracting beta-band oscillations using a modified periodogram-Welch method (Figure 3). Prior to post-analysis, standard preprocessing was applied to the EEG data. Data acquisition was performed using a Neuroscan software-based system (https://compumedicsneuroscan.com/).

These oscillations, subsequently converted into beta-band phase dynamical series via Phase Locking Value (PLV), served as input for the analysis. Multi-regional phase space analysis, in conjunction with four machine learning algorithms (Gradient Boosting, Random Forest, k-Nearest Neighbors, and Decision Tree), was then applied to predict working memory (WM) states from these phase series.

2.5.1. EEG Data Preprocessing

This study utilized a 32-channel Ag/AgCl electrode cap, designed in accordance with the international 10-20 system standards for EEG signal acquisition. A Neuroscan software-based data acquisition system was employed, with a raw signal sampling rate of 1024 Hz, which was subsequently down-sampled to 128 Hz. During the analog signal processing stage, a 0.5-40 Hz band-pass filter was applied. The data processing workflow adhered to standardized pre-processing protocols:

Electrode Positioning and Channel Selection: Verified the spatial positions of 32 valid electrodes based on the international 10-20 system and removed redundant electrode channels that were not in use.

Sampling Rate Verification: Confirmed that the raw data sampling rate was strictly maintained at 1024Hz, was consequently down-sampled to 128 Hz, in accordance with the technical specifications of the equipment.

Reference Potential Correction: Applied the whole-head average reference method to perform spatial re-referencing of the raw signals.

Digital Filtering: Implemented zero-phase FIR filtering in the MATLAB/EEGLAB environment to maintain the effective frequency band of 0.5-40 Hz.

Artifact Removal: Identified and removed physiological interference components, such as eye movements and muscle activity, using the Independent Component Analysis (ICA) algorithm.

Data Cleansing: Identified and deleted data segments containing extreme amplitudes (>±100 μV) or abnormal noise using a semi-automatic detection method.

2.5.2. Modified Periodogram -Welch Method for PSD Calculation

The traditional periodogram method involves directly performing a Fourier transform on the signal and calculating the power spectrum. However, this approach is notorious for its high variance and unsmooth spectral estimation, particularly when dealing with limited signal lengths, leading to issues such as spectral leakage and insufficient frequency resolution. To address these limitations, the Welch method employs a segment averaging strategy that utilizes overlapping segments to reduce variance while maintaining an optimal trade-off between bias and variance. This method has become the most prevalent non-parametric technique for power spectral density (PSD) estimation in brainwave analysis due to its improved statistical performance. In this study, we implemented the Welch PSD estimation using the PWELCH function in MATLAB, applying it to brainwave data recorded from 32 channels [27]. The Welch power spectral density estimation method for brainwave signals $Xn$ can be formally described as follows [28]:

Segmentation with Overlap: The brainwave signal sequence is divided into overlapping segments. Let the length of each segment be $L$. Starting from the second segment ($i$=1), the first D sample values of the $i$-th segment overlap with the last D sample values of the $(i-1)$-th segment. The $i$-th segment can be mathematically represented as:

$$x_i = \begin{cases} x_N[i(L-D)+n], n = 0,1...L-1, i = 0,1,...K-1 \\ 0 \end{cases} \quad (2)$$

The relationship between the total sample length N, the overlap D, and the segment length L is given by: $N = L+(L-D)(K-1)$, where $K = \frac{N-L}{L-D}+1$.

Windowing and the Fourier Transform: Each segment is multiplied by a smoothing window ($w(n)$) before computing the Fourier transform.

Averaging Power Spectral Density: The squared magnitudes of the Fourier transform of all segments are averaged, and amplitude compensation is applied to compute the power spectral density estimate using the modified periodogram method:

$$S_x(e^{jw}) = \frac{1}{K}\sum_{i=1}^{K}\frac{1}{LU}\left|x_i(e^{jw})\right|^2 \quad (3)$$

Here, $U = \frac{1}{L}\sum_{n=0}^{L-1}w(n)^2$ represents the average power of the window function.

2.5.3. Brain Topography Map Method

In the previous step, we calculated the PSD for 32 channels. Following this, we focused on the beta frequency band and determined the power of each channel within this band by integrating and summing the PSD. Finally, we applied the min-max normalization method to convert the band power into a dimensionless index within the 0-1 range, thereby providing standardized node features for subsequent analysis of EEG signal characteristics. The mathematical expressions are as follows:

$$BandPower = \sum\nolimits_{f_{low}}^{f_{high}} PSD(f) \qquad (4)$$

$$band\_ratio = \frac{(band\_power - \min(band\_power))}{(\max(band\_power) - \min(band\_power))} \qquad (5)$$

Where $[f_{low}, f_{high}]$ represents the target frequency band, and *band_ratio* denotes the Normalized Node Activation Degree (*NNAD*), which is the normalized value of a node's activation.

Further, *NNAD* values for each node were calculated using Eq. (5). These NNAD values were then projected onto a 2D brain map, adhering to standard anatomical coordinates, using color gradients to represent cortical node activation values. Mapping analyses were conducted based on the MNE-Python Package (https://mne.tools/).

2.5.4. Dynamical Phase Synchronization Analysis

Phase synchronization analysis is a powerful mathematical tool in the field of neural oscillation synchronization, capable of precisely describing synchronized activities between brain regions. In this study, we utilize a commonly used indicator derived from the phase locking method—the Phase Locking Value, *PLV* [29].

For two continuous-time signals (*x(t)*) and (*y(t)*), the PLV is used to characterize the synchronization between the two signals:

$$PLV = \left\| \exp(i\{\varphi_x(t) - \varphi_y(t)\}) \right\| \qquad (6)$$

In the equation above, φx(t) and φy(t) denote the instantaneous phases of x(t) and y(t), respectively. The phase is determined through the Hilbert transform, as described below:

$$x(t) = \frac{1}{\pi} p.v \int_{-\infty}^{+\infty} \frac{x(a)}{t-a} da \qquad (7)$$

Here, "p.v." represents the Cauchy principal value (PV). The instantaneous phase is subsequently determined by:

$$\varphi_x(t) = \arctan\frac{x(t)}{x(t)} \qquad (8)$$

Let N be the length of the synchronously acquired, discrete, real-valued signal data for both channels. Using the formulas provided above, compute the instantaneous phases of the two signals, denoted as φ1(n) and φ2(n), respectively. The instantaneous phase difference between the two channels is:

$$\theta(n) = \varphi_1(n) - \varphi_2(n) \qquad (9)$$

The phase coherence index R is defined as:

$$R = \left| \frac{1}{N} \sum_{j=0}^{N-1} e^{i\theta(n)} \right| \qquad (10)$$

As indicated by the formula, the theoretical range of R is confined to the closed interval [0,1]. When R=0, it signifies that the instantaneous phase difference between the two neural oscillatory signals is uniformly distributed in a random manner, corresponding to a state of complete phase desynchronization. Conversely, when R=1, the two signals display a stable phase difference, indicating phase synchronization.

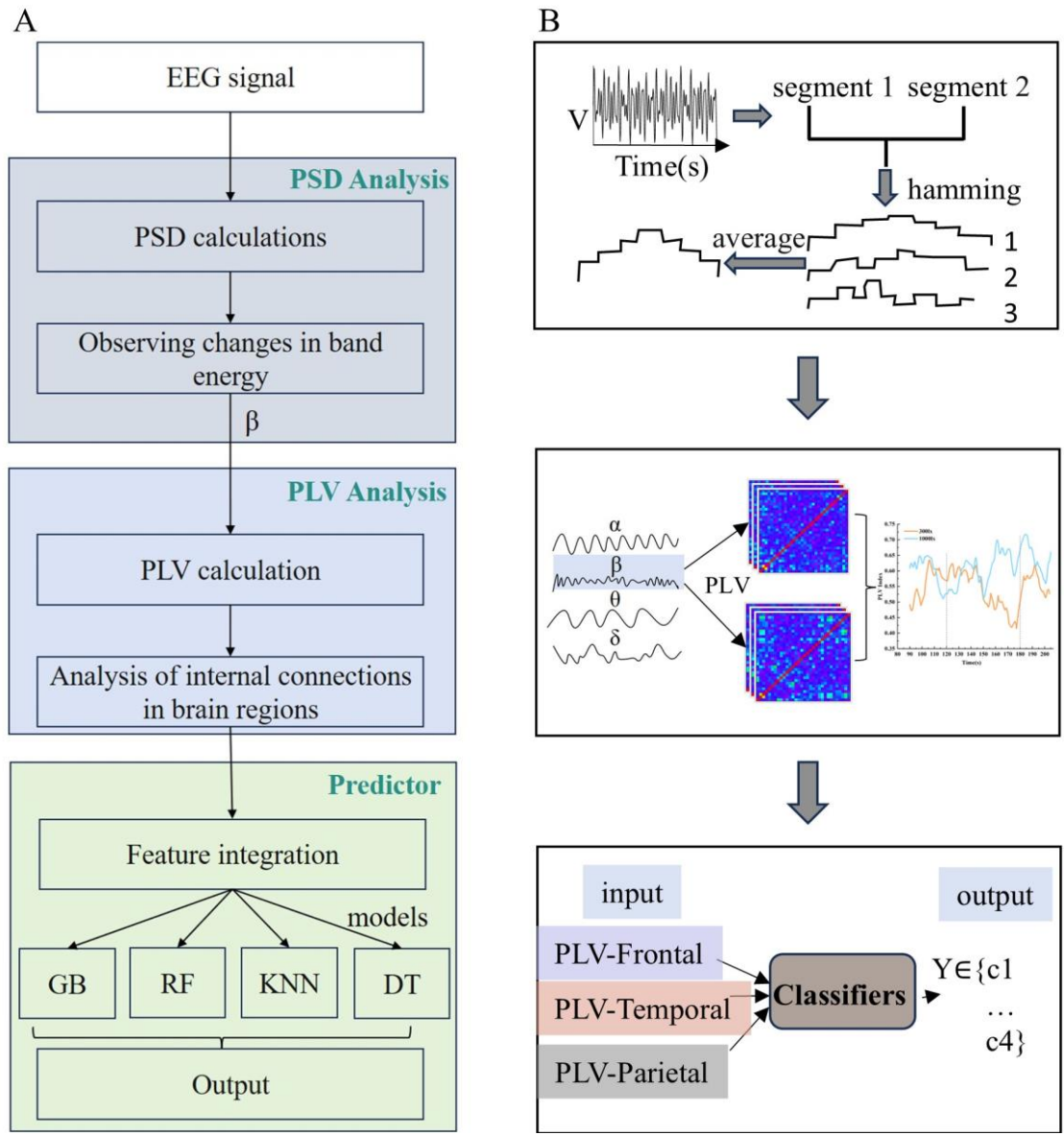

**Figure 3.** Algorithm flowchart of the beta-based phase synchronization analysis and machine-learning classification. A. Algorithm modules in this paper include PSD analysis, PLV analysis, and machine learning. B. Detailed schematic diagram for the corresponding left algorithm modules.

## 3. Results

*3.1. Dynamical Phase Synchronization Patterns in Beta Band during four WM Brain States*

As shown in Figure 4A, the beta-band brain topography map indicates that the prefrontal lobe remains highly active during the memory task. In contrast, the intensity of brain activity in the temporal lobe decreases with changes in memory activity and light environment (i.e., state1-state4). In the comparative analysis under different lighting conditions, it was found that the prefrontal lobe and central connections, as well as the parietal lobe, were more active under 1000 lx. As light intensity increases, subjects' performance in the working memory task improves, and the brain's neural activity patterns shift from a dual-stable state (temporal and prefrontal lobes) to a triple-stable state (temporal, prefrontal, and parietal lobes). This change suggests that appropriate lighting conditions can effectively promote the rational allocation and efficient utilization of brain neural resources, thereby enhancing working memory efficiency.

Figure 4B presents the statistical results of node activation values across the prefrontal, temporal, and parietal regions for 22 subjects. The data reveal that the prefrontal lobe maintains consistently high activation levels across all four states, with the highest activation observed at 1000 lx. In contrast, the temporal lobe exhibits higher activation values at 300 lx, which decline progressively as illumination levels increase. Conversely, the parietal lobe initially shows lower activation, which gradually increases in response to changing lighting conditions. These findings suggest that for the different of environmental and task conditions, the prefrontal region consistently dominates the center-control role, while the temporal and the parietal regions potentially act the coordinating and balancing role in controlling the adaptability of brain WM system.

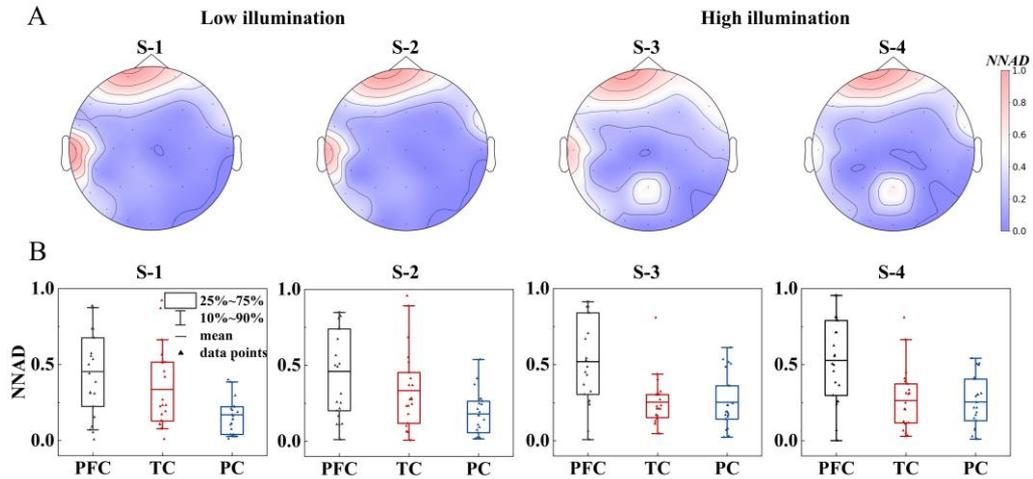

**Figure 4.** Mapping and statistics of Beta-Band Brain Topography. A. Beta-band activity topography via node-power-derived NNAD, visualizing beta activation across two cognitive experiments and lighting conditions. B. Stats on activation values of key nodes in specific brain regions (22 subjects): prefrontal (black, nodes: FP1, FP2); temporal (red, nodes: FT9, T7); parietal (blue, nodes: P3, Pz). state 1(S-1):300lx-recall, state 2(S-2):300lx-sequence memory, state 3(S-3):1000lx-recall, state 4(S-4):1000lx-sequence memory.

In this study, the PLV was employed to analyze dynamic patterns of averaged phase synchronization in the beta-band frequency range across three distinct brain regions: PFC, TC, and PC (denoted as $\bar{R}_x(t)$, where x represents PFC, TC, or TC), the temporal courses was obtained by introducing sliding windows. The beta-band phase information was specifically selected due to its established association with working memory (WM) states, as supported by previous studies [30]. For this analysis, we selected specific electrode sites corresponding to each brain region: FP1 and FP2 for the PFC, FT9 and T7 for the TC, and P3 and Pz for the PC. he experimental results, as illustrated in Figure 5, demonstrated distinct neural coordination patterns across the four experimental conditions. Throughout the entire time course, the PFC exhibited consistently high levels of internal neural coordination, suggesting its central executive role in WM processes [10]. In contrast, the PC displayed a significant upward trend in internal neural coordination, while the TC showed a corresponding downward trend. These findings imply a potential compensatory or balancing mechanism between the PC and TC regions, which is further elaborated in the subsequent discussion and analyses presented in Figure 6 and Figure 7.

Figure 5 identified a fundamental transition in neural synchronization dominance across the four WM states. Under dim light conditions (300lx), temporal cortex internal synchronization prevailed, facilitating efficient information transmission via robust neural signal coordination among temporal cortical regions. Conversely, with increased illumination (1000lx), parietal cortex internal coordination emerged as the predominant characteristic. This finding, corroborated by beta-power 2D topological patterns (Figure 4), underscores a light-dependent modulation of WM state dynamics, highlighting the emergent role of parietal cortex coordination in brighter environments.

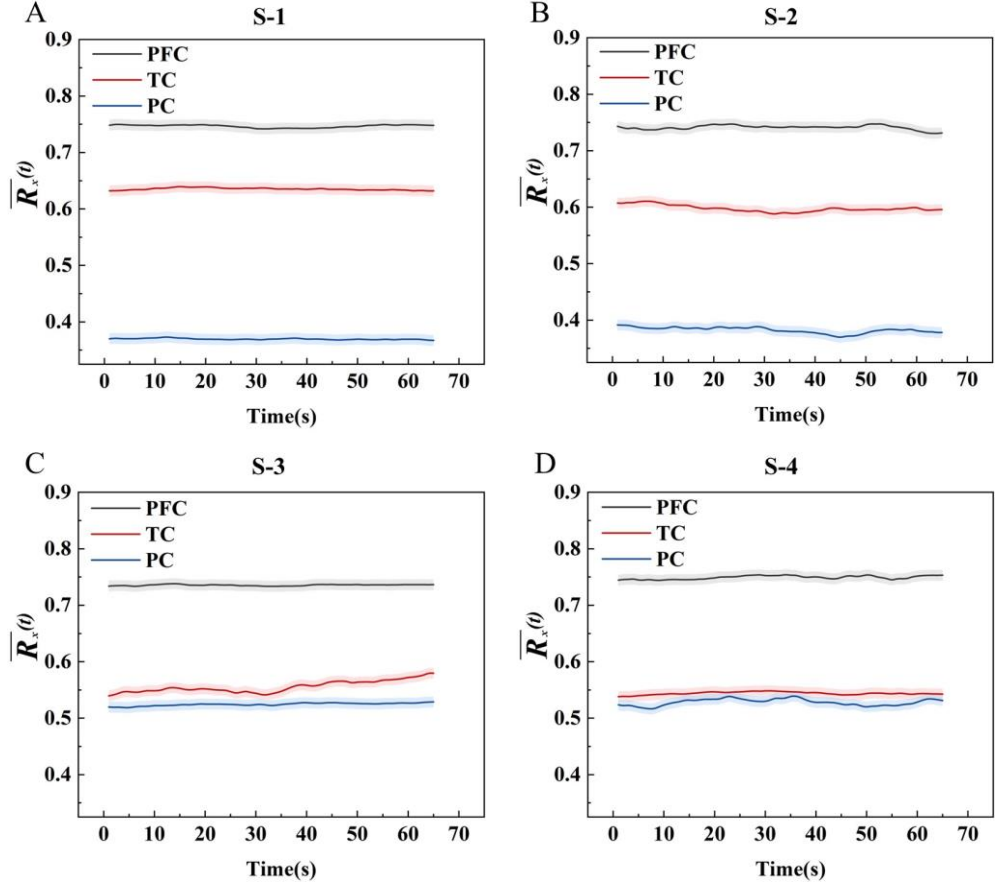

**Figure 5.** Dynamical patterns of local synchronization factor ($\bar{R}_x(t)$) in TC, PC, and PFC regions for the four WM Brain States from state 1（S-1） to state 4(S-4), respectively.

*3.2. Phase Trajectory Patterns of Multi-Regional Synchronization Dynamics*

To investigate the functional interactions among the PFC, TC, and PC in maintaining distinct WM states, we performed a dynamic $\bar{R}_x(t)$-based Phase spaces analysis across both three-dimensional and two-dimensional phase spaces. As shown in Figure 6A, the phase-space clusters corresponding to S-3 and S-4 (brighter condition) exhibit significantly higher values along the PFC dimension compared to clusters representing S-1 and S-2 (less bright condition). Given the improved behavioral performance associated with the brighter condition in Figure 2D and E, this finding further underscores the central executive role of the PFC in WM functions. The phase-space results, as presented in Figure 6B and C, did not reveal significant interregional interactions between the PFC and PC or between the PFC and TC. However, Figure 6D reveals a distinct trend across task states: TC activity progressively decreases while PC activity increases, approaching equilibrium along the y = x balance line. This equilibrium point, as will be discussed in subsequent sections, serves as a critical indicator of the multi-regional neural patterns underlying WM state plasticity.

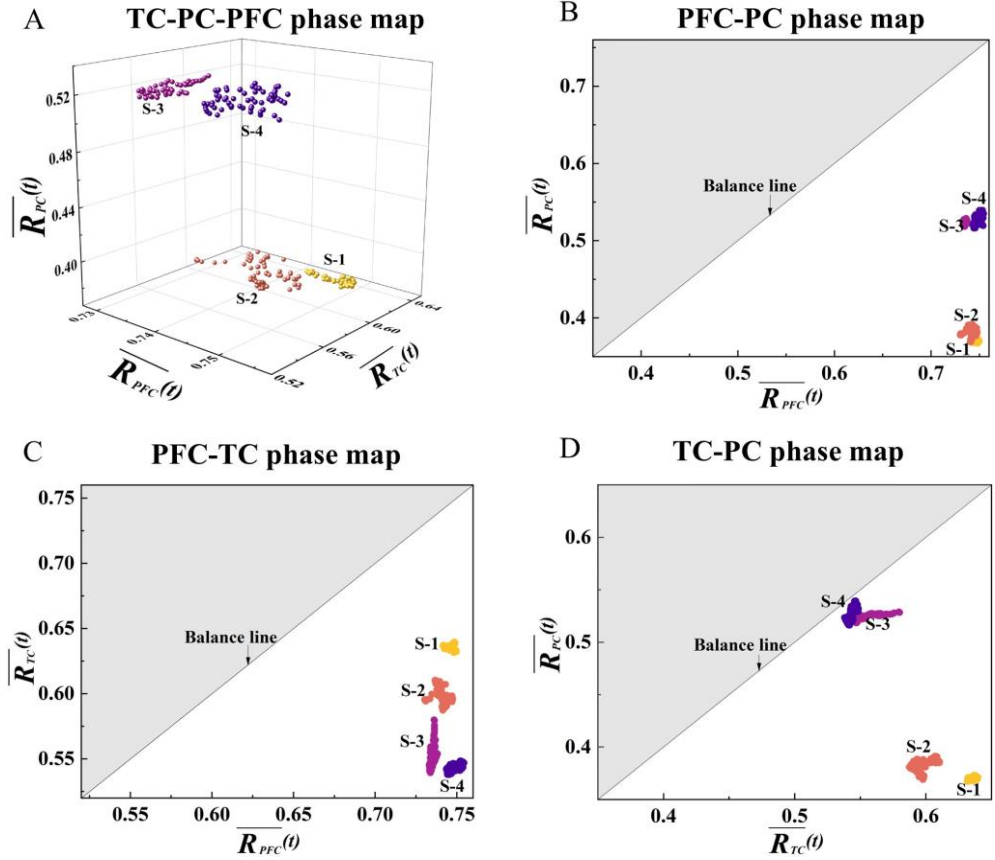

**Figure 6.** Phase trajectory patterns of multi-regional synchronization factor $\bar{R}_x(t)$ in 3D TC-PC-PFC space (A), 2D PFC-PC space (B), PFC-TC space (C), and TC-PC space (D), respectively.

*3.3. Recognizing WM States through Dynamical Temporal-Parietal Balanced Patterns*

To provide a more rigorous and clear interpretation of the equilibrium state under discussion, this study adopts a dual analytical framework comprising both geometric and arithmetic approaches. Figure 7A presents a visual representation of the distances between the centroids (x0, y0) of the data distributions across different states and the equilibrium lines, thereby illustrating the temporal evolution of data centroid convergence toward the equilibrium lines through a sequence of four states. In a complementary fashion, Figure 7C provides a quantitative assessment based on the regional differences in $\bar{R}_x(t)$ values between the TC and PC regions.

Specifically, the $\bar{R}_{TC}(t)$ sequence is mathematically defined as f(t), while the $\bar{R}_{PC}(t)$ sequence is represented as g(t). A time-varying difference sequence, h(t), is subsequently constructed by calculating the TC-PC differences. The proximity of h(t) to zero indicates a more balanced state of neural synchronization between the TC and PC regions. The observed results, as shown in Figure 7B and Figure 7D, demonstrate a progressive reduction in these inter-regional differences across the four-state sequence, with the values approaching zero. The convergence of findings from both analytical methods is corroborated by the brain topography map (Figure 4), thereby providing robust evidence that TC-PC equilibrium effectively optimizes the allocation of neural memory resources.

To systematically evaluate the predictive capabilities of $\bar{R}_x(t)$ metrics across the temporal, parietal, and prefrontal cortices in characterizing WM states, this study employed a multi-model machine learning framework. Specifically, four distinct classification algorithms—gradient boosting, random forest, k-nearest neighbors, and decision trees—were applied to assess predictive performance, as detailed in Figure 8A. The analysis utilized the temporal averaged synchronization factor $_{<\bar{R}_x(t)>}$ derived from 16 participants across the three brain regions as input features, with corresponding WM brain states designated as output labels. To comprehensively examine the predictive efficacy of different regional combinations, we established four comparative feature sets: TC-PC, TC-PFC, PC-PFC, and TC-PC-PFC. Classification performance was systematically evaluated across these feature combinations.

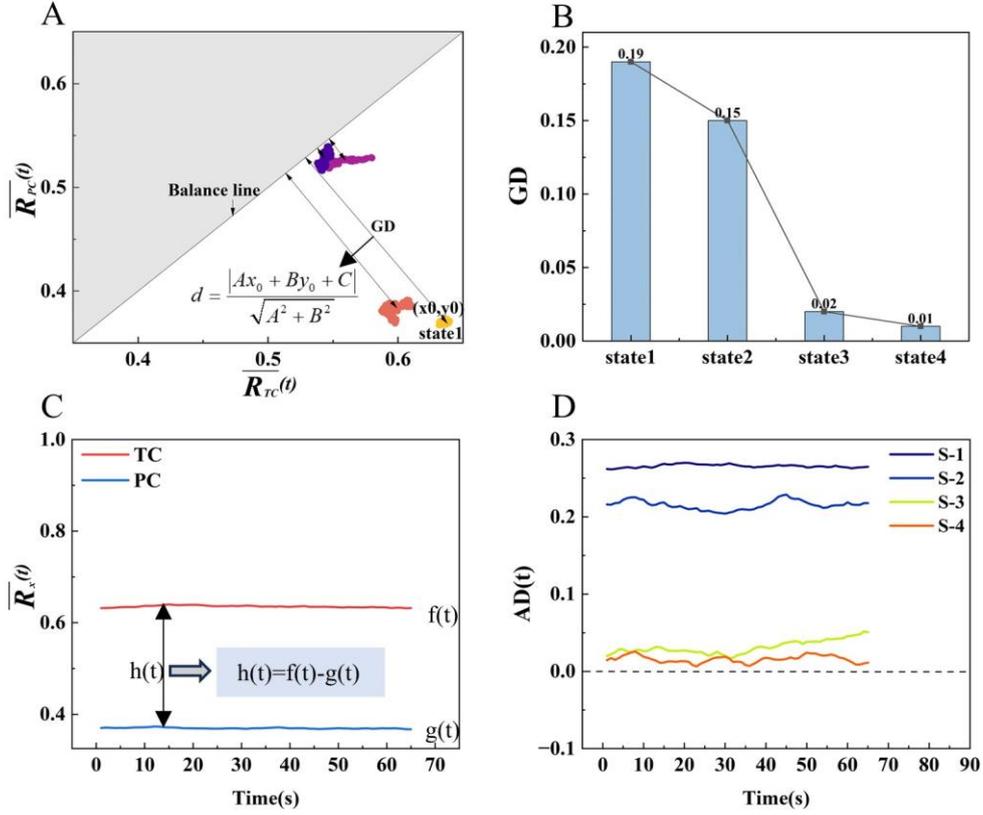

**Figure 7.** Parietal-Temporal synchronization balance and the working memory adaptability. A. The algorithms of measuring the geometric distance (GD) from each state's TC-PC phase-map distribution centroid (x0, y0) to the balance line. B. The histograms of GD for the four WM states. C. The algorithms of measuring the arithmetic distance (AD) from each state's dynamical difference between TC synchronization factor ($\bar{R}_{TC}(t)$) and PC synchronization factor ($\bar{R}_{PC}(t)$). D. The dynamical AD(t) trends for the four WM S-1 (purple), S-2 (blue), S-3 (green) and S-4 (red), respectively.

As illustrated in Figure 8B, the TC-PC combination demonstrated the highest classification accuracy across all four machine learning methodologies. Among all models, the KNN model demonstrates the best classification performance, with a classification accuracy rate of 95%. This finding underscores the robust generalizability of the TC-PC pattern in effectively categorizing samples into their respective WM states, maintaining consistent performance even when different machine learning algorithms were applied. These results highlight the superior discriminative power of combined TC-PC metrics in classifying WM brain states, providing empirical evidence for the significant correlation between TC-PC balance and WM state differentiation.

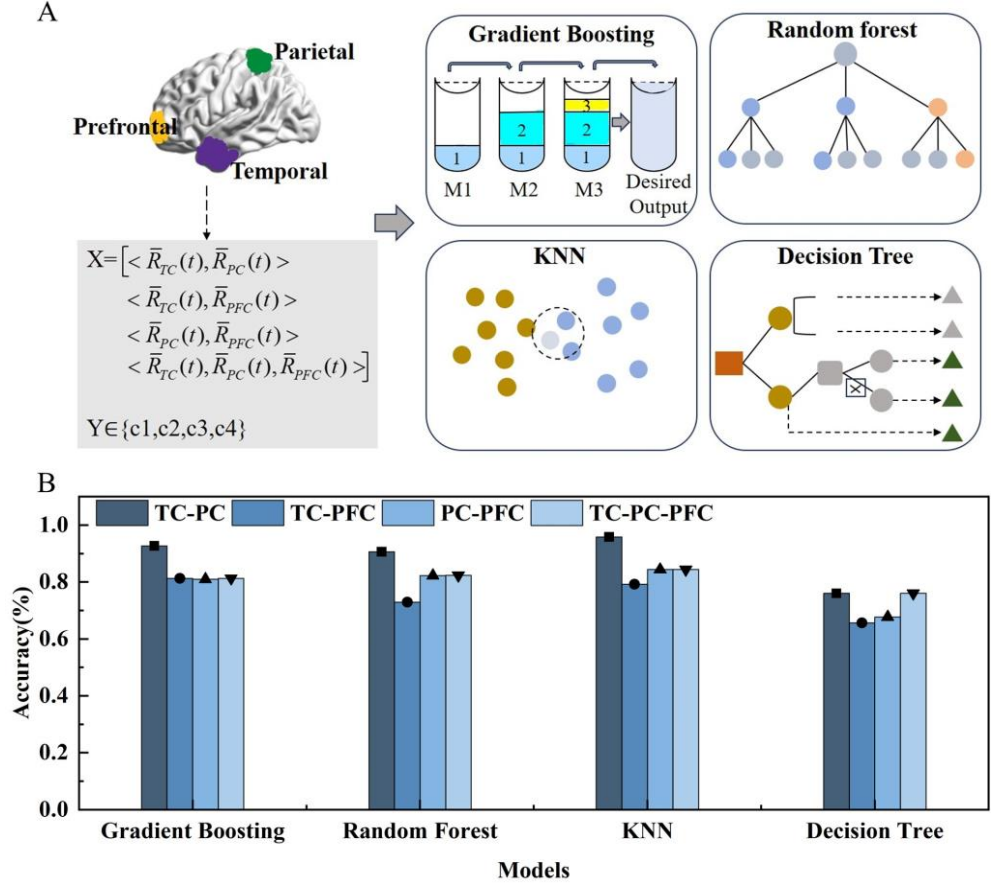

**Figure 8.** Assessment of the predictive performance of the temporal averaged synchronization factor ($<\bar{R}_x(t)>$) in the TC, PC, and PFC using machine learning techniques. A. Supervised models (Gradient Boosting, Random Forest, KNN, Decision Tree) used $<\bar{R}_x(t)>$ from TC, PC, PFC as features (brain - region combinations: TC - PC, TC - PFC, PC - PFC, TC - PC - PFC) to predict WM brain states (c1 - c4). B. Prediction accuracy of the four models across brain - region combinations.

## 4. Discussion

Our findings demonstrate that the prefrontal cortex maintains a consistent beta - rhythm local synchronous state regardless of changes in working memory load or lighting conditions. These results provide robust evidence supporting the established view proposed by Baddeley and Hitch [31,32] regarding the prefrontal cortex's critical role in regulating and coordinating working memory operations. Its local synchronicity remains dynamically stable across varying memory conditions, likely reflecting the prefrontal cortex's homeostatic regulation necessity to coordinate bottom - up attentional processes [33,34] and top - down memory decision - making processes [35,36] under diverse cognitive demands and environmental conditions.

Our analysis also reveals dynamic synchronization-balanced patterns of the temporal - parietal network through varying lighting and memory load conditions. Our findings suggest this balance may be a critical whole - brain network steady - state condition for enhancing working memory, as seen in improved task accuracy and better memory search capacity. These results extend prior studies focusing solely on the temporal or parietal cortex in working memory [37,38]. Consistent with research advocating a whole - brain systems approach to understanding working memory [39], our study highlights the importance of especially temporal - parietal interplay. Statistical results (Figure 7) and machine learning predictions (Figure 8) show this balance's significance in working memory state pattern recognition, refining our understanding of the whole - brain dynamic mechanisms underlying working memory.

Working memory theory posits that its operation relies on a distributed brain network centered on the prefrontal cortex (PFC), incorporating subcortical structures such as the hippocampus and thalamus, along with coordinated activity across cortical regions including the temporal and parietal cortices. Central to this framework is the PFC's role in top-down control, particularly in regulating memory storage and retrieval processes. While previous studies have extensively examined the contributions of the thalamus and hippocampus to memory retrieval, the mechanisms by which the PFC integrates information across distributed cortical networks to facilitate retrieval remain understudied. Our findings provide novel evidence from a network dynamics perspective, demonstrating that the PFC adaptively modulates beta-band synchronization with the aid of temporal and parietal regions in response to varying environmental demands and task requirements. This discovery offers a critical extension to the central executive theory of working memory.

## 5. Conclusions

Our study investigates the dynamic adaptability of working memory, leveraging a novel framework that integrates EEG-based multi-regional beta phase synchrony dynamics with machine learning. Key findings reveal that, unlike the prefrontal cortex—whose local beta rhythm synchrony remains stable across varying memory loads and lighting conditions—the temporal-parietal region exhibits pronounced dynamic adjustments: temporal-parietal phase clustering outperforms other regional combinations in distinguishing memory states, and illumination-enhanced environments optimize its synchronization balance (quantified via proposed arithmetic and geometric balance indices). Machine learning validates that temporal-parietal synchrony serves as the dominant feature for cross-task working memory state classification, confirming its superior applicability in decoding and predicting such states.

These discoveries deepen our understanding of working memory mechanisms by highlighting the critical role of temporal-parietal balance, extending beyond traditional central executive models. Practically, they contribute a precise prediction algorithm that facilitates the development of a low-dimensional system utilizing temporal and parietal EEG channels—with direct value for real-time cognitive ergonomics assessment in BCIs and optimized human-machine interaction.

**Author Contributions:** Jiajia Li: Writing – review & editing, Writing – original draft, Visualization, Validation, Resources, Methodology, Investigation, Funding acquisition, Formal analysis, Data curation, Conceptualization; Tian Guo : Writing – review & editing, Writing – original draft, Visualization, Validation, Software, Investigation, Formal analysis, Data curation; Fan Li: Writing – review & editing, Visualization, Validation, Software; Jian Song : Writing – review & editing, Writing – original draft, Visualization, Validation, Supervision, Resources, Project administration, Methodology, Investigation, Funding acquisition, Conceptualization.

**Funding:** This work was supported by the National Natural Science Foundation of China (Grant Nos. 81870863,12002251), the Postdoctoral Scientific Research Foundation, General Hospital of Central Theater Command (Program No. 20220224KY29).

**Institutional Review Board Statement:** This study was conducted in accordance with the Declaration of Helsinki and has been approved by the Ethics Committee of the Central Theater General Hospital of the Chinese People's Liberation Army (Reference No.: [2020]041-1).

**Informed Consent Statement:** Informed consent was obtained from all subjects involved in the study.

**Data Availability Statement:** The raw data supporting the conclusions of this article will be made available by the authors on request. The data are not publicly available due to privacy.

**Acknowledgments:** This work was supported by the National Natural Science Foundation of China (Grant Nos. 12572068, 81870863), the Postdoctoral Scientific Research Foundation, General Hospital of Central Theater Command (Program No. 20220224KY29).

**Conflicts of Interest:** The authors declare that they have no known competing financial interests or personal relationships that could have appeared to influence the work reported in this paper.